\documentclass[twocolumn,letterpaper]{igs}

\usepackage{igsnatbib}
\usepackage{stfloats}
\usepackage{igsupmath}

\ifx\pdftexversion\undefined
\usepackage[dvips]{graphicx}
\else
\usepackage[pdftex]{graphicx}
\usepackage{float}
\restylefloat{table}
\restylefloat{figure}
\fi

\usepackage{color}

\begin{document}

\title[Radio-frequency attenuation at Summit Station]{An \emph{in situ} measurement of the radio-frequency 
attenuation in ice at Summit Station, Greenland}

\author[Avva and others]{Jessica AVVA$^1$,
  John M. KOVAC$^2$, Christian MIKI$^3$, David SALTZBERG$^4$, Abigail~G.~VIEREGG$^1$}

\affiliation{%
$^1$Department of Physics, Enrico Fermi Institute, Kavli Institute for Cosmological Physics, 
University of Chicago,
Chicago, IL 60637, USA\\
E-mail: javva@uchicago.edu, avieregg@kicp.uchicago.edu\\
$^2$Harvard-Smithsonian Center for Astrophysics, Cambridge, MA 02360, USA\\
$^3$Department of Physics, University of Hawaii at Manoa, Honolulu, HI 96822, USA\\
$^4$Department of Physics and Astronomy, University of California, Los Angeles, Los Angeles, CA 90095, USA}
\abstract{We report an \emph{in situ} measurement of the electric field attenuation length $L_\alpha$
at radio frequencies for the bulk ice at Summit Station, Greenland, made 
by broadcasting radio-frequency signals vertically through the ice and measuring the relative power in the
return ground bounce signal.  
We find the depth-averaged field attenuation length to be \mbox{$\langle L_\alpha \rangle =947^{+92}_{-85}$~m} 
at 75~MHz. 
While this measurement has clear radioglaciological applications, the radio 
clarity of the ice also has implications for the detection of ultra-high energy (UHE) astrophysical 
particles via their radio
emission in dielectric media such as ice.
Assuming a reliable extrapolation to higher frequencies,
the measured attenuation length at Summit Station 
is comparable to previously measured radio-frequency attenuation lengths 
at candidate particle detector sites around the world, and strengthens the case
for Summit Station as a promising northern site for UHE neutrino detection.}

\maketitle

\section{Introduction} 
We report a measurement of the radio-frequency electric field attenuation length in the ice at the Summit 
Station site in Greenland.  Our interest in the radio properties of glacial ice stems from the 
applications to particle astrophysics, but these measurements are also of interest for the development 
of radar systems that probe sub-surface features in glacial ice, such as ice strata and sub-glacial 
lakes and streams.

We are ultimately interested in building a detector to 
search for radio emission created when the highest energy astrophysical neutrinos interact in a large 
volume of a dielectric material~\citep{rice,aura,anitaInstrument,ara,arianna}.
Glaciers are the leading candidate medium for a detector for ultra-high energy (UHE) neutrinos 
because the cold ice temperature leads to a 
long radio attenuation length~\citep{bogorodsky} and they have large volume.  These 
two properties combine to allow for a large enough instrumented detector volume to
have the sensitivity required to detect significant numbers of UHE neutrinos, which are very rare~\citep{halzen}.

When a neutrino interacts with a dielectric material, such
as glacial ice, it initiates a shower of charged particles 
approximately 0.1~m in diameter and tens of meters long.
The charged particles in the shower move 
faster than the speed of light in the medium, which is reduced compared to 
the speed of light in a vacuum by the index of refraction of ice, $n$.  
This causes Cerenkov radiation, the electromagnetic 
analogue to a sonic boom.  Because the size of the particle 
shower is small compared to the wavelength of radio 
waves, the radio component is emitted coherently at frequencies up to a few~GHz,
and for high-energy showers is the strongest component of the radiation~\citep{askaryan}.
This coherence effect was later confirmed in the laboratory 
in a variety of media, including ice, 
using showers initiated by high-energy electron and photon beams~\citep{saltzberg, anitaAskaryan}.

\begin{figure}[ht!]
  \centering{\includegraphics[width=0.45\textwidth]{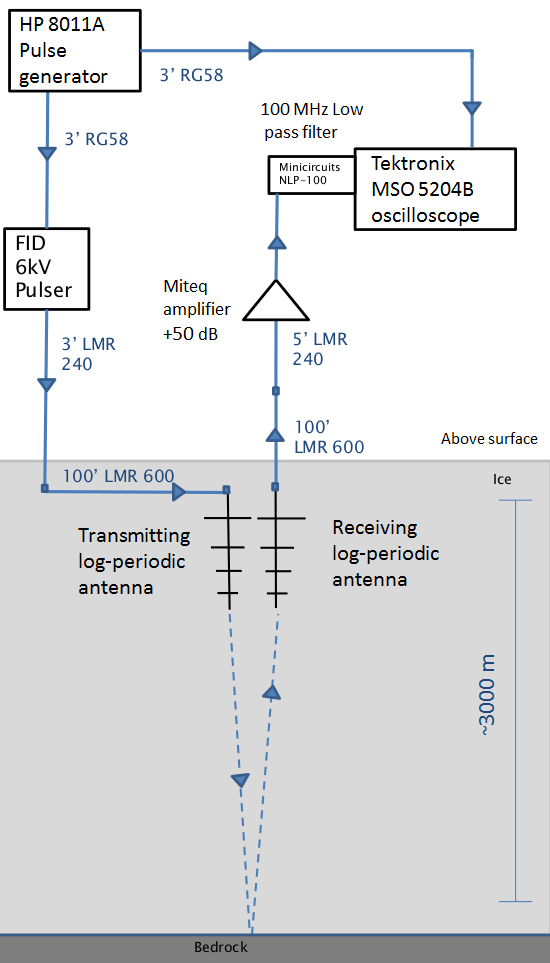}}
  \caption{A drawing of the experimental setup for the ground bounce with low-frequency antennas.  
    We used a FID Technologies 6~kV high voltage pulser, 
    triggered by a Hewlett Packard 8011A pulse generator to transmit
    through a log-periodic antenna.  We used a second log-periodic antenna
    coupled to a $+50$~dB Miteq amplifier to boost the 
    received signal and a Tektronix MSO5204B to record the data.}
  \label{fig:sketch}
\end{figure} 

The distance the radio-frequency signals can propagate depends on the properties of the 
glacial ice at the specific site; the attenuation length 
varies from site to site by large factors. 
Attenuation length corresponds to 
the rate of neutrino detection, so \emph{in situ} measurements at specific sites are essential. 
Previous measurements have been made of the radio properties of the ice at 
sites in Antarctica (Taylor Dome, the South Pole, and the Ross Ice Shelf) 
to determine the best southern sites for UHE neutrino 
detection~\citep{barwick,besson,barrella,ara,newArianna},
and have found a depth-averaged field attenuation length $\langle L_\alpha \rangle$ at 300~MHz of
$1660^{+255}_{-120}$~m over the top 1500~m of ice at 
the South Pole~\citep{ara}.

There have been previous measurements in the frequency range of interest (hundreds of MHz) 
at Summit Station that have investigated layering 
in the ice~\citep{paden} for their glaciologic implications, and the radar attenuation length has recently 
been measured at many sites across the Greenland Ice Sheet~\citep{macgregor}.
In this paper, we report an \emph{in situ} measurement of the radio attenuation length of the bulk glacial 
ice at the Summit Station site made using similar methods to previous measurements at other 
sites~\citep{barwick,besson,barrella,ara,newArianna}.

\section{Experimental Approach} 
Our approach is similar to previous work in our field~\citep{barrella, besson, barwick}.  From the surface, we
transmit a high-voltage impulse with broadband frequency content into the ice and measure the power
in the return signal as a function of frequency with a second, identical antenna.  We compare this ground
bounce to the transmission of the same impulse through a short distance in air, 
directly from one antenna to the other.  After accounting for the geometric factor $1/r$ (for electric
field strength), where $r$ is the propagation distance,
the remaining loss in electric field 
in the ground bounce return pulse is attributed to attenuation and scattering in the ice,
which goes as $\mathrm{exp}(-r/\langle L_\alpha \rangle)$.
We use this technique of comparing the through-air data to the in-ice data, rather than calculating
the expected power in the return ground bounce signal from the known power transmitted by the system, 
to reduce systematic uncertainty.

We adopt a similar notation and technique 
to~\cite{barrella}.  We define $P_\nu$ to be the power spectral density at
frequency $\nu$ of the impulse as received by a $50~\Omega$ receiver, and $V_\nu$ to 
be $\sqrt{P_\nu \times 50~\Omega}$.
The quantity 
$V_{\nu,\mathrm{ice}}$ is $V_\nu$ measured after passing through the ice, bouncing off of the bedrock, and 
traveling back
to the receiver over a total distance $d_\mathrm{ice}$, and $V_{\nu,\mathrm{air}}$ is $V_\nu$ after 
being transmitted directly through the air between the two antennas separated by $d_\mathrm{air}$. 

Therefore,
\begin{equation}
V_{\nu,\mathrm{ice}}/V_{\nu,\mathrm{air}} = (d_\mathrm{air} / d_\mathrm{ice}) e^{-d_\mathrm{ice}/\langle L_\alpha \rangle},
\end{equation}
where $\langle L_\alpha \rangle$ is the depth-averaged field attenuation length over the entire depth of the ice. 

Solving for $\langle L_\alpha \rangle$ gives
\begin{equation} 
\langle L_\alpha \rangle = d_\mathrm{ice}/ \ln(\frac{V_{\nu,\mathrm{air}}~d_\mathrm{air}}{V_{\nu,\mathrm{ice}}~d_\mathrm{ice}}).
\end{equation}

The ratio of the power transmitted by the antenna into ice compared to air, $T_\mathrm{ratio}$, 
due to a small difference in coupling between the antenna and ice compared to air, affects the measured
attenuation length $\langle L_\alpha \rangle$.  
The assumed power reflection coefficient $R$ at the ice-bedrock interface
also affects the measured $\langle L_\alpha \rangle$.
Including both of these effects, $\langle L_\alpha \rangle$ becomes
\begin{equation} 
\langle L_\alpha \rangle = d_\mathrm{ice}/ \ln(T_\mathrm{ratio}~\sqrt{R}~\frac{V_{\nu,\mathrm{air}}~d_\mathrm{air}}{V_{\nu,\mathrm{ice}}~d_\mathrm{ice}}).
\label{eqn:attenLength}
\end{equation}
In the above equation, we 
have included a factor of $\sqrt{T_\mathrm{ratio}}$ for the relative transmission of electric field for 
each antenna (the transmitter and the receiver), yielding a total factor of $T_\mathrm{ratio}$.

\begin{figure*}[ht]
  \centering{\includegraphics[width=18cm]{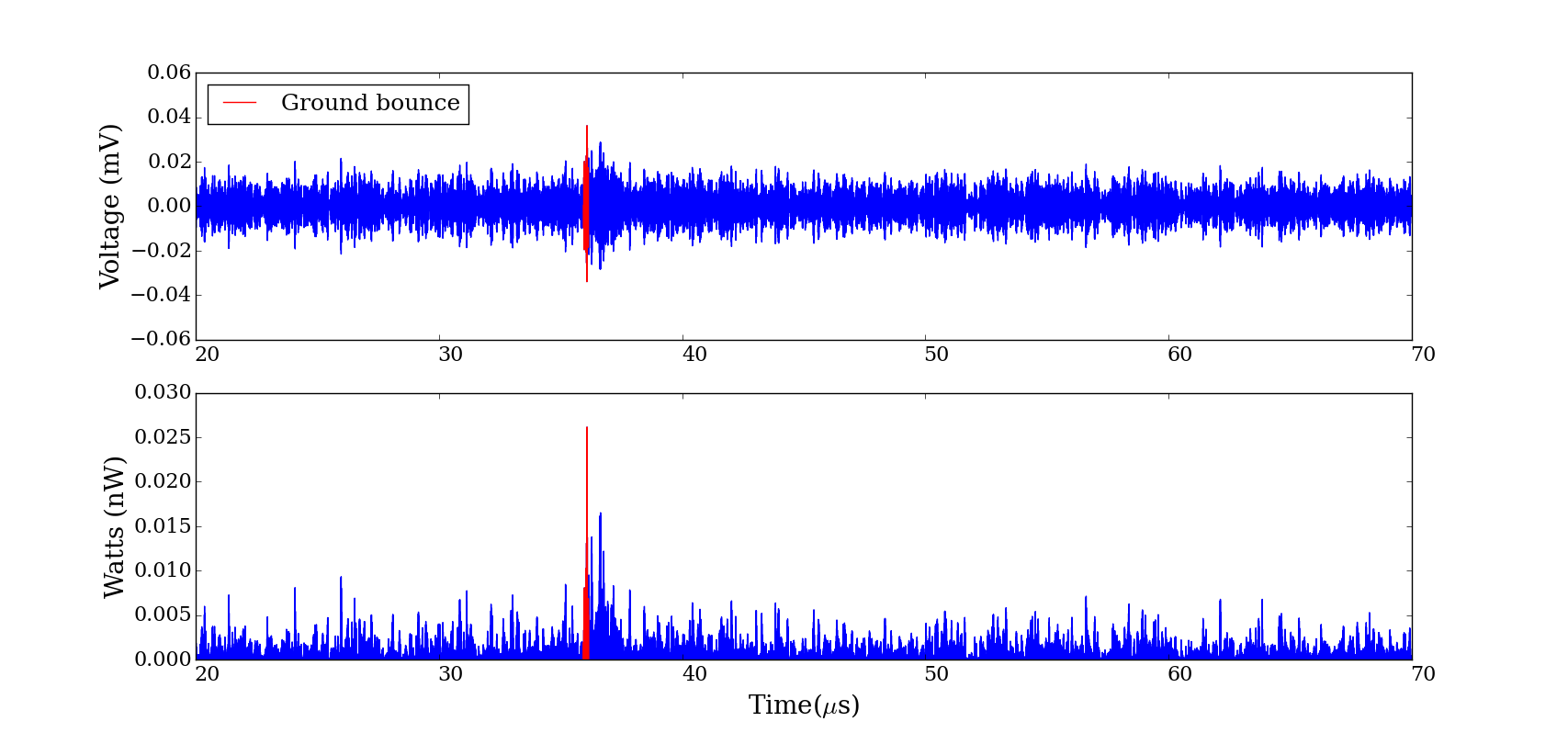}}
  \caption{Top: The received voltage in a 50~$\Omega$ receiver as a function of time 
    with the log-periodic antennas buried in the snow, pointed downward.  A large reflection 
    is evident at 36.1~$\mu$s (highlighted in red), consistent with a depth of $3014^{+48}_{-50}$~m. 
    Bottom: The power received as a function of time, derived from the top panel.  The red region denotes
    the 200~ns wide time window used for analysis.}
  \label{fig:twopulsesprelim}
\end{figure*}

\subsection{Experimental Setup}

\begin{figure}
  \centering{\includegraphics[width=9cm]{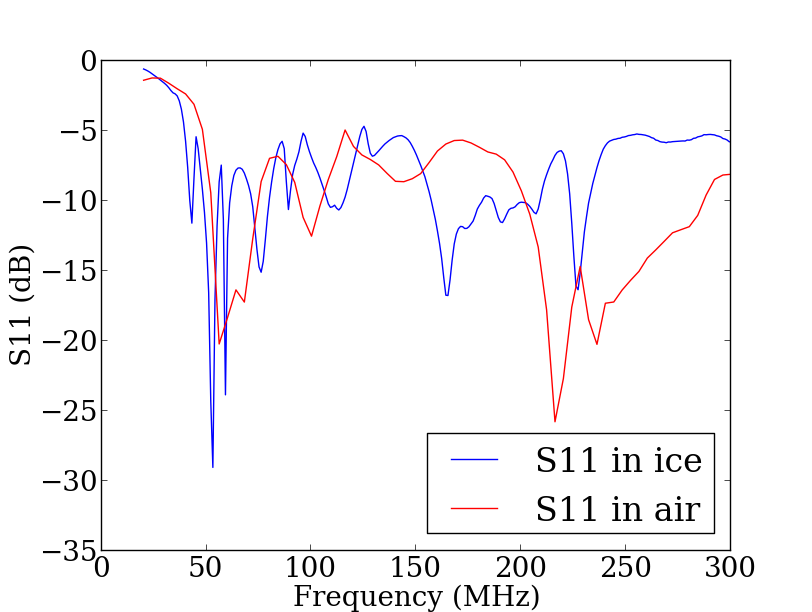}}
  \caption{The return loss ($\mathrm{S}_{11}$) of the log-periodic antenna used for the ground bounce measurement for
    the antenna buried below the surface and packed with snow (blue), compared to the same
    antenna in air (red).}
  \label{fig:s11s}
\end{figure} 

The measurement that we report here was performed in June 2013 at the Summit Station site,
and is a ground bounce, described above
and shown in a sketch in Figure~\ref{fig:sketch}.  The measurements were made
at N 72$^\circ$ 37' 20.7'' W 38$^\circ$ 27' 34.7'', 
near Summit Station, the highest point on the Greenland ice sheet. 
We transmitted a fast, high-voltage impulsive signal, generated by a FID Technologies 6~kV high-voltage 
pulser\footnote{http://www.fidtechnologies.com}.  The high-voltage pulser was 
triggered by a Hewlett Packard 8011A pulse generator, which also triggered the oscilloscope that 
recorded the received signal.  We transmitted the high-voltage impulse through 30~m of 
LMR-600 cable\footnote{http://www.timesmicrowave.com} before sending it out of a high gain antenna aimed down
toward the bottom of the glacier.  We received the signal with an identical antenna, also aimed downward,
46~m away along the surface of the snow.  
We then amplified the received signal with a 
Miteq\footnote{http://www.miteq.com} amplifier ($+50$~dB of gain, model AFS3-00200120-10-1P-4-L) 
before passing it through a 100~MHz low-pass filter and reading it out with a Tektronix MSO5204B 
oscilloscope that was set to average over 10,000 impulses.  We included the 100~MHz low-pass filter   
in the system to reduce intermittent noise pickup from man-made sources.
The antennas were buried in the snow
and packed with snow to ensure the best coupling between the antenna and the snow, reducing 
Fresnel effects at the snow-air interface.  

To record the signal that we transmitted through the system, we also took data with the antennas aimed toward
each other through the air, 46~m apart.  This normalization run through air had the same 
cabling and setup as the ground bounce measurement, but with an additional 40~dB attenuator on the input
to the Miteq amplifier to avoid saturating the amplifier with the large signal.  

To reduce the 
effect of reflections off of the surface of the snow in the normalization run through air,
we took data with the antennas 2~m above the surface of the snow.  We tested that
the effect of reflections off of the surface of the ice on the received signal is small by varying
the height of the antennas and looking for changes in the observed signal and saw none.  We also 
used these antennas previously for testing in a high bay in a similar configuration, 
and placed radio frequency absorber along the surface of the floor and observed no significant 
change to the signal.

We used a pair of log-periodic antennas with good transmission between 60~and~100~MHz and 
$\sim6$~dBi of gain\footnote{http://www.scannermaster.com}.  
The return loss ($\mathrm{S}_{11}$) of the log-periodic antenna when buried below the surface and 
packed with snow compared to the antenna in air is shown in Figure~\ref{fig:s11s}.  
The transmission band of the antenna moves down slightly in frequency when coupled to the snow,
evidenced by the change in the -3~dB point of the antenna and 
expected from the higher index of refraction of snow compared to air.
There is good transmission
in the frequency range used in analysis (65-85~MHz) in air and when coupled to the firn.

\subsection{Data Analysis}

The top panel of Figure~\ref{fig:twopulsesprelim} shows a long trace recorded with the low-frequency antennas 
pointed downward and buried in the snow.  The bottom panel of the figure is the power in the return signal 
as a function of time, derived from the top panel. 
The trace has been filtered to 65-85~MHz using a Butterworth 
filter of order three to limit the noise contribution out of the band of the
system (defined at the low frequency end by the antenna transmission 
coefficient and the high frequency end by the low-pass filter) and
extract a clear signal.  There is a clear reflection at an absolute time 
of 36.1~$\mu$s, which is consistent with the timing expected for a signal that bounced off of the 
ice-bedrock interface at a depth known from GRIP borehole measurements~\citep{grip}.  
From the absolute time of 36.1~$\mu$s, accounting for the 
measured system delay of 430~ns, we can measure the total round-trip 
distance through the ice, $d_{\mathrm{ice}}$, using the
relationship

\begin{equation} 
d_{ice} = c_n \Delta t,
\label{eqn:depthtiming}
\end{equation}
where $c_n$ is the speed of light in the medium and $\Delta t$ is the total time of flight
through the ice.  $c_n$ is related to the dielectric constant of the medium via

\begin{equation} 
c_n = \frac{c}{\sqrt{\epsilon '}}, 
\end{equation}
where $c$ is the speed of light in a vacuum and $\epsilon '$ 
is the real component of the complex dielectric constant of the material. The 
index of refraction of the medium, $n$, is equal to $\sqrt{\epsilon '}$.
We use a firn model based on the 
measured density profile,~$\rho$,~at Summit Station from~\cite{arth}
to determine the index of refraction as a function of depth for the firn layer.
Our firn model has two regimes: one that describes the density from the surface to 
$15$~m below the surface, and a second that describes the firn from $15$~m
to $100$~m below the surface.  By $100$~m, the firn has transitioned to 
glacial ice.  For radio propagation in 
glacial ice, the dielectric constant is related to the density of the ice~\citep{kovacs} by 

\begin{equation}
\epsilon ' = (1+0.845 \rho)^2,
\end{equation}
allowing us to calculate $n$, and subsequently $c_n$, as a function of depth for the firn.
We assume an index of refraction of glacial bulk ice of $1.78 \pm 0.03$ for ice below
$100$~m deep~\citep{bogorodsky}.
We calculate that the depth of the observed ground bounce is $3014^{+48}_{-50}$~m, which leads
to $d_{\mathrm{ice}}=6028^{+96}_{-100}$~m.  This is consistent with the known depth of 
the ice at Summit Station from GRIP borehole measurements~\citep{grip}.

We note that there is a second and smaller
reflection $0.6$~$\mu$s after this initial ground bounce (corresponding to 91~m farther).  
In principle, it is valid to choose any return signal, as long
as we include the correct distance to the bounce in our calculations and make a realistic assumption about
the power reflection coefficient at the interface, $R$.  In practice, we ran our 
data analysis over each of the two return impulses (36.1~$\mu$s and 36.7~$\mu$s), 
and there is little difference in the extracted attenuation length value.

\begin{figure}[ht]
  \centering{\includegraphics[width=9cm]{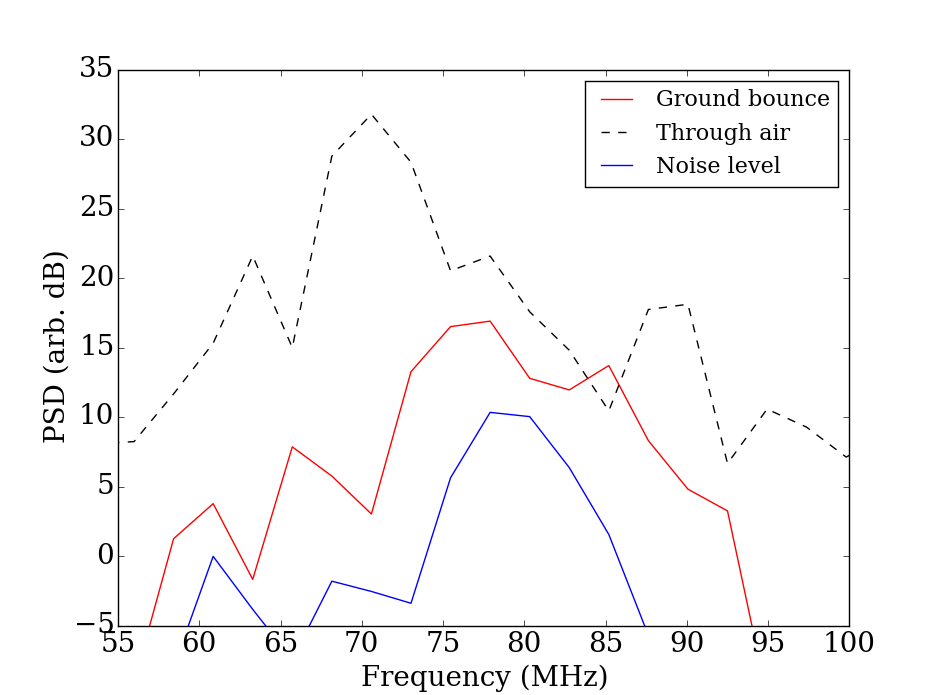}}
  \caption{The power spectral density of the received ground bounce signal compared to the noise level 
    in the trace. The power spectral density of the through-air normalization signal, with 40~dB
      of additional attenuation compared to the ground bounce signal, is
      shown for comparison with a dashed line.  All data have been filtered (65-85~MHz), 
        time-windowed, and zero-padded 
        in the same way (see the text for more details).}
  \label{fig:signalandnoise}
\end{figure} 

Figure~\ref{fig:signalandnoise} shows the time-windowed power spectral density of the ground bounce
in red, time-windowed on the red region (200~ns) in Figure~\ref{fig:twopulsesprelim}, filtered to 65-85~MHz using a
Butterworth filter of order three, and zero-padded in the 
time domain.  The blue line in Figure~\ref{fig:signalandnoise} shows the typical noise level in the trace
taken from a typical (noise-only) time window well after the observed ground bounce in time. It has been
time-windowed, filtered, and zero-padded in the same way as the ground bounce signal region. 

\begin{figure}[h]
  \centering{\includegraphics[width=9cm]{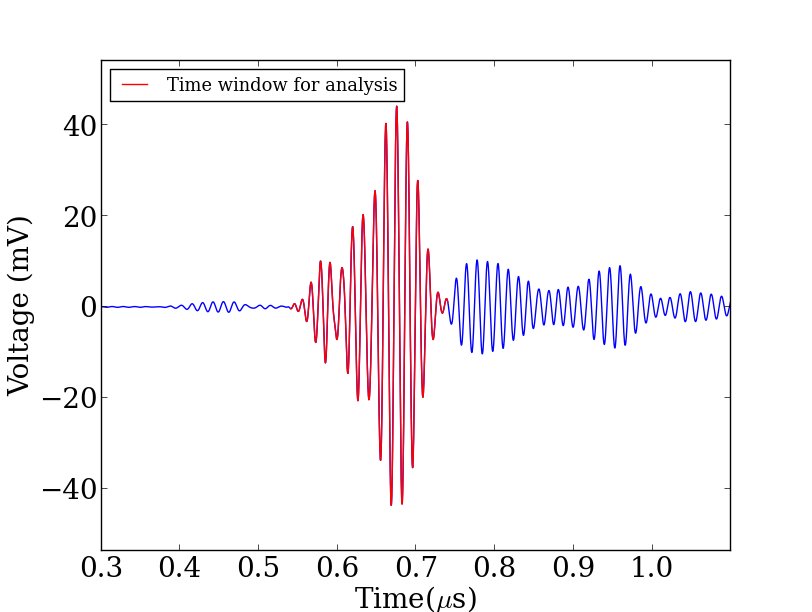}}
  \caption{The transmitted impulsive signal received directly through 46~m of air, Butterworth 
    filtered to 65-85~MHz.  The system used is the same as for the ground bounce, but with 40~dB 
    of additional attenuation on the receiver.  The red region shows the time window used for analysis (200~ns wide).}
  \label{fig:airpulse}
\end{figure}

To measure the radio-frequency loss observed through the ice, we compare the power in the 
received ground bounce signal to the power in the signal measured through the air with the same setup
but with the antennas 46~m apart and aimed directly toward each other.  We use this method
because the response of our system cancels out in the normalization between the ground bounce
and the direct through-air signal.
Figure~\ref{fig:airpulse}
shows the waveform of the normalization signal transmitted directly through the air, filtered
to 65-85~MHz in the same way as the ground bounce signal, and with an additional
40~dB attenuator inserted before the low-noise amplifier compared to the sketch in Figure~\ref{fig:sketch}.
The power
spectral density for the through-air normalization signal, with 40~dB of additional attenuation compared to
the ground bounce signal, is shown with a dashed line in 
Figure~\ref{fig:signalandnoise}.  It has been time-windowed, filtered, and zero-padded in the same way as the 
ground bounce signal region.

From the ground bounce signal power, 
we subtract the power spectral density of the typical noise region shown in Figure~\ref{fig:signalandnoise}. 
We process the through-air signal in the same way as
the ground bounce and noise traces, producing a power spectral density of the through-air signal
that is time-windowed, filtered, and zero-padded in the same way as the ground bounce.  
We then integrate the power between 65-85~MHz in the noise-subtracted ground bounce and the through-air
signal.  Using the square root of the total integrated power in this frequency range for $V_{ice}$ and 
$V_{air}$ and accounting for the 40~dB of additional attenuation on the receiver, we calculate the 
depth-averaged field attenuation length through the ice using 
Equation~\ref{eqn:attenLength}.  We assume
the power reflection coefficient $R$ to be 0.3, which is typical of the ice-bedrock 
interface~\citep{barwick}, and the antenna 
transmission ratio $T_\mathrm{ratio}$ to be 1.0, which was the measured value at 75~MHz for similar
antennas by~\cite{barrella} on the Ross Ice Shelf.  
We calculate that $\langle L_\alpha \rangle =947$~m at 75~MHz.

\subsection{Systematic Errors}
{\bf Index of Refraction:} One error on the attenuation length measurement comes from the 
uncertainty
in the assumed index of refraction of ice, which translates linearly to an uncertainty on the distance
to the ground bounce.  The index of refraction of glacial ice has been measured to be
temperature independent in the measured temperature range of the ice at Summit Station
with $n=1.78\pm0.03$ by~\cite{bogorodsky}, corresponding 
to an uncertainty on the depth of the ice of $3014^{+48}_{-50}$~m and on 
$\langle L_{\alpha} \rangle$ of $^{+18}_{-18}$~m.  

{\bf Antenna Coupling:} The transmission coefficient ($\mathrm{S}_{21}$) of the antennas changes 
when the antennas are coupled to the
ice compared to when they are coupled to the air.  
We made measurements of the reflection coefficient ($\mathrm{S}_{11}$) 
for these antennas in the field (shown in Figure~\ref{fig:s11s}).  Assuming that all power that 
is not reflected is transmitted, the antennas transmit over 98\% of the power 
in the frequency range of interest for the analysis (65-85~MHz) both when coupled to ice and coupled to air,
indicating that $T_\mathrm{ratio}=1.0$.
We also use direct measurements of the transmission properties of similar antennas made by~\cite{barrella}, 
which indicate a value of $T_\mathrm{ratio}=1.0$ at 75~MHz, 
with an uncertainty of about 10\%. 
We include this 10\% uncertainty in the power transmitted
due to the different coupling to the air and snow of the log-periodic antennas ($T_\mathrm{ratio}$).
This corresponds to an uncertainty on $\langle L_\alpha \rangle$ 
of $^{+14}_{-16}$~m.

{\bf Power Reflection Coefficient:} The power reflection coefficient at the ice-rock interface is not well known.
In our calculation, we assume a power reflection coefficient of 0.3, which is typical of a 
bedrock-ice interface.  Table~\ref{tab:reflection} shows the effect on field attenuation length
for different assumptions on the power reflection coefficient.  Assuming a perfect reflector at the bottom ($R=1$)
is a conservative assumption, and yields a shorter attenuation length.  We take the uncertainty on
$\langle L_\alpha \rangle$ to be $^{+89}_{-82}$~m: 
the range of field attenuation lengths derived using a power reflection coefficient of $R=0.1$ to $R=1.0$.

\begin{table}[ht]
\begin{center}
\begin{tabular}[c]{|c|c|}
\hline \hline
Power Reflection Coefficient & Field Attenuation Length $\langle L_\alpha \rangle$ \\
\hline
0.1 & 1038~m\\ 
0.3 & 947~m \\
1.0 & 865~m\\ 
1.25 & 853~m\\
\hline\hline
\end{tabular}
\end{center}
\caption[Power Reflection Coefficients]{The depth-averaged 
electric field attenuation length as a function of the choice of power reflection coefficient
  at the ice-bedrock interface.  We include a calculation with $R$=1.25, an extremely pessimistic
  case that would require magnification effects at the ice-bedrock interface.
 }\label{tab:reflection}
\end{table} 

{\bf Other Possible Sources of Error:}
Other possible sources of uncertainty include uncertainty in the density profile of the firn,
the effect of birefringence in the ice,
the effect of any physical bedrock features, and any change in gain
of the antennas when coupled directly to the snow.  The effect of the first on our measurement is small.  

Birefringence has been shown in general to cause losses as large as 10~dB in our frequency range, but it is 
suspected
that the loss due to birefringence is much smaller at places along an ice sheet divide
such as the Summit Station site~\citep{paden}.  This is due to the lack of strong horizontal 
preference in the ice crystal fabric because of slow ice flow 
and is evidenced by previous measurements at the site~\citep{paden}.  We plan to make further measurements
of the effect of birefringence on the measured attenuation length at the Summit Station site.

Features at the bedrock surface could serve to either magnify or demagnify the reflected signal, depending
on the geometry of the surface.   Our calculation assumes that the signal is reflected off of a 
flat, horizontal surface and suffers no magnification effects.  This is a good approximation of the
ice-bedrock layer around the Summit Station site, which does not have dramatic features at the 
bedrock surface~\citep{bamber}.

Since the frequency response of the antennas shifts lower in frequency by $\sim20\%$ 
when the antennas are placed in the snow (see Figure~\ref{fig:s11s}), the beam pattern at a given frequency
in snow corresponds to a beam pattern at a $\sim20\%$ higher frequency in air. For log-periodic 
antennas, we do not expect the gain to be dramatically different over the frequency range of interest, so
the contribution to the total error is subdominant.

{\bf Total Error:} 
Combining all of the quantifiable uncertainties (antenna coupling, index of refraction, and reflection 
coefficient), we find a the total uncertainty on $\langle L_\alpha \rangle$ of $^{+92}_{-85}$~m:

\subsection{Results}
We have calculated the depth-averaged electric field attenuation length at 75~MHz through an
analysis of the ground bounce measurement, and have quantified the systematic errors associated
with the measurement.  We find the depth-averaged attenuation length including all quantifiable uncertainties
to be $\langle L_\alpha \rangle =947^{+92}_{-85}$~m at 75~MHz.

\section{Discussion and Summary}  

\begin{figure}[ht]
  \centering{\includegraphics[width=9cm]{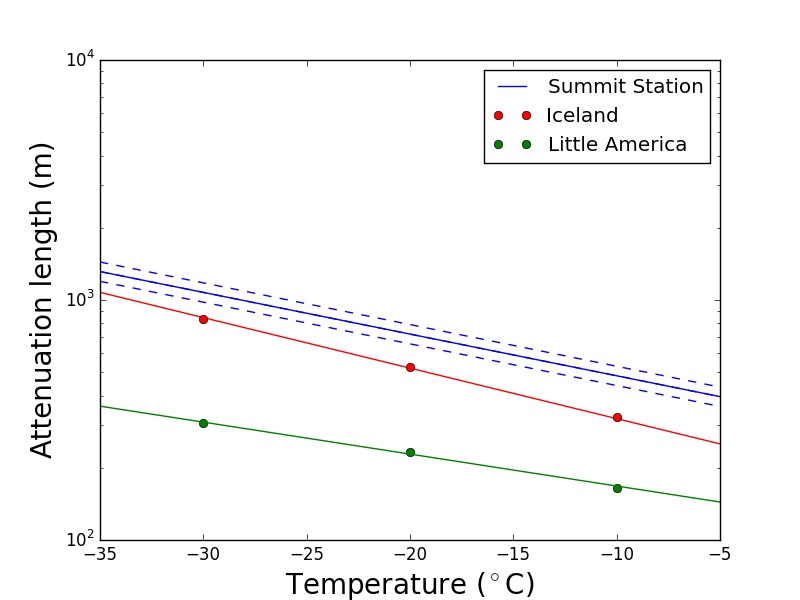}}
  \caption{The electric field attenuation length as a function of temperature for the Summit Station site,
    shown with the blue line.  We have 
    assumed that the relationship between attenuation length and temperature is consistent with
    the measured attenuation length vs. 
    temperature from~\cite{bogorodsky}, also shown on this plot (red and green lines).  
    The dashed lines denote
    $\pm1\sigma$.}
  \label{fig:tempattenfit}
\end{figure}

\begin{figure}[ht]
  \centering{\includegraphics[width=9cm]{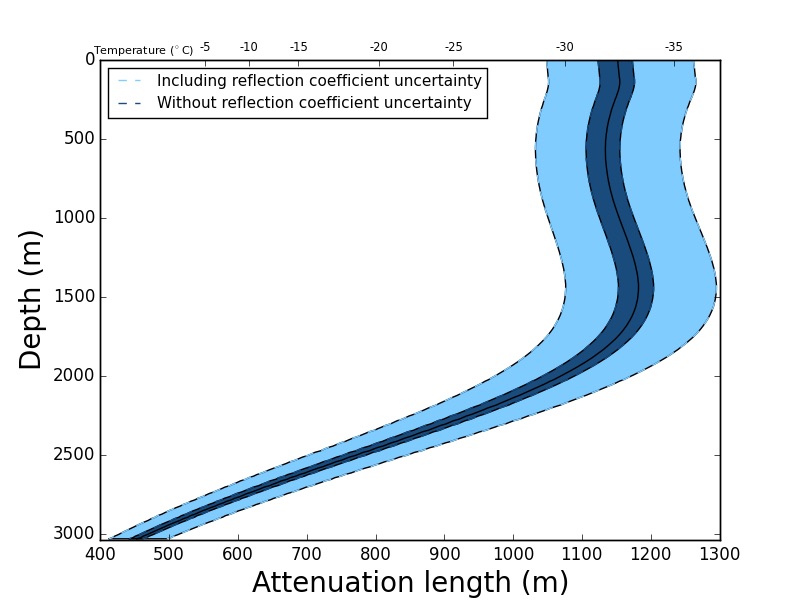}}
  \caption{The extracted electric field 
    attenuation length profile as a function of depth for the Summit Station site at 75~MHz.  
    We have combined our measurements with 
    the temperature profile measured for the GRIP borehole~\citep{grip,boreholetemp} 
    and the measured attenuation length
    vs. temperature from~\cite{bogorodsky}. The shaded region denotes $\pm1\sigma$.  We have not included 
    a firn density correction, which would be a small effect.}
  \label{fig:depthprofile}
\end{figure}

We combine our measurement of the depth-averaged field attenuation length $\langle L_\alpha \rangle$ with 
the measured temperature
profile of the ice at the Summit Station 
Site from the GRIP borehole~\citep{grip,boreholetemp} and the measured dependence of 
attenuation length on temperature from~\cite{bogorodsky} 
(shown in Figure~\ref{fig:tempattenfit}) to extract a profile of attenuation
length as a function of depth.  For the dependence of attenuation length on temperature, 
we assume the average slope of the two locations on which~\cite{bogorodsky} report,
shown as the blue line in Figure~\ref{fig:tempattenfit}. 
The field attenuation length as a function of depth is shown in Figure~\ref{fig:depthprofile}.
The average field attenuation length over the upper 1500~m, from where the vast majority of the neutrino events
that a surface or sub-surface radio detector could detect would originate, 
is $\langle L_\alpha \rangle$=$1149^{+112}_{-103}$~m at 75~MHz.

To compare with measurements made at the South Pole and on the Ross Ice Shelf, we
extrapolate our results to 300~MHz~\citep{besson,barwick,barrella}.  This is a model-dependent
extrapolation.   We use an ensemble of 
measurements of the attenuation length as a function of frequency of glacial
ice from Antarctica, Iceland, and Greenland~\citep{bogorodsky, walford} 
and (Westphal cited in~\citep{westphal}) 
to perform a linear extrapolation.  We take the average of the best 
and worst cases from this ensemble, yielding a linear extrapolation with a slope of $-0.55$~m/MHz.  
This yields an estimate of the field attenuation length in the top 1500~m at 300~MHz of $1022^{+230}_{-253}$~m.
We note that this extrapolation introduces a large 
uncertainty (reflected in the error bars quoted) because the properties of glacial ice at
different locations are variable, and a direct measurement at Summit Station at 
higher frequencies would be more robust.  We take the additional uncertainty to be the scatter 
in the measurements of attenuation length as a function of frequency.

We also took data in a similar experimental configuration
with Seavey broadband quad-ridged horn antennas from Antenna Research 
Associates\footnote{www.ara-inc.com} that are sensitive between 200-1200~MHz.
The experimental setup that we used for these higher frequencies did not have enough sensitivity
make a direct measurement at 300~MHz due to a smaller antenna
effective area and less power in the broadband high-voltage pulser.  
However, as a consistency check, we can place an upper limit
on the depth-averaged attenuation length at 300~MHz from the higher-frequency data.  Following the
same procedure outlined previously, but at 300~MHz, we calculate that our system was 
sensitive to a depth-averaged attenuation length of 1100~m or longer at 300~MHz.  Applying the frequency
extrapolation described in this Section to the measured depth-averaged attenuation length
\mbox{$\langle L_\alpha \rangle$ of $947^{+92}_{-85}$~m} at 75~MHz yields a depth-averaged attenutation
length at 300~MHz of $823^{+189}_{-209}$~m, consistent with our directly-derived upper limit.

We compare the results of our measurement and extrapolation
of the field attenuation length at Summit Station in the top 1500~m at 300~MHz of $1022^{+230}_{-253}$~m
with other \emph{in situ} measurements of radio 
attenuation at possible sites for neutrino detectors.  For deep sites, we follow the convention
established by~\cite{ara} and compare the attenuation length
in the top 1500~m of the ice, since that is where the vast majority of neutrino events that a surface 
or sub-surface detector occur.  For the site
on the Ross Ice Shelf, we use the depth-averaged attenuation length, since the total depth of the ice
is much less than 1500~m.
At 300 MHz, the radio attenuation length measured on the Ross Ice Shelf~\citep{barrella} is 
$\langle L_\alpha \rangle =411$~m with an experimental uncertainty of about 40~m
averaged over all depths for the 578~m thick ice.  This measurement has been redone recently,
with a consistent result~\citep{newArianna}.  The attenuation length at the South Pole has been measured
to be $\langle L_\alpha \rangle =1660^{+255}_{-120}$~m for the top 1500~m~\citep{ara}.  The longer attenuation length
at the South Pole in the upper 1500~m compared to Summit Station 
can mainly be attributed to the fact that the ice 
is colder in the upper 1500~m
($\sim-50^\circ$C at the South Pole compared to $\sim-30^\circ$C at the Summit Station site).

Our measurement of radio-frequency 
attenuation length at Summit Station is 25\% longer
than the comparison
that we can make to the previously measured South Pole attenuation length~\citep{ara}, 
after accounting for the difference in temperature profiles 
between the South Pole and Summit Station and assuming the
relationship between attenuation length and temperature as described above and from~\cite{bogorodsky}.
This is not meant to be an exact prediction, but rather a comparison of the ice after removing obvious
differences such as temperature and depth.
Our measurement is also consistent with recent radar measurements of attenuation length at 195~MHz across
the Greenland Ice Sheet, including near Summit Station~\citep{macgregor}.

The Summit Station site is an appealing choice to host a detector for UHE neutrinos.  
The site sits on top of the deepest part of the roughly 3~km deep Greenland ice sheet, 
providing a huge detection volume.
The radio attenuation length of the ice is comparable to sites in the Antarctic.
The development of a northern site for UHE neutrino detection will allow for different 
sky coverage compared to developing Antarctic arrays~\citep{ara,arianna}.

\section{Acknowledgements}
We would like to thank CH2MHill and the US National Science 
Foundation (NSF) for the dedicated, knowledgeable, and extremely helpful logistical support team, 
particularly to K. Gorham.
We are deeply indebted to those who dedicate their careers to 
help make our science possible in such remote environments.  
This work was supported by the NSF's Office of Polar Programs
(PLR-1103553), the US Department of Energy, 
the Kavli Institute for Cosmological Physics at the 
University of Chicago, the University of California, Los Angeles, 
and the Illinois Space Grant Consortium.
We would like to thank P. Gorham for support of C. Miki.
We would also like to thank Warner Brothers Studios for lending us parkas for the expedition.

\bibliography{attenuationPaper}
\bibliographystyle{igs}  

\end{document}